\input harvmac
%\draftmode
\def\journal#1&#2(#3){\unskip, \sl #1\ \bf #2 \rm(19#3) }
\def\andjournal#1&#2(#3){\sl #1~\bf #2 \rm (19#3) }

\def\ie{{\it i.e.}}
\def\eg{{\it e.g.}}

\def\frac#1#2{{#1\over#2}}

\def\half{\frac12}

\def\inbar{\,\vrule height1.5ex width.4pt depth0pt}
\def\IC{\relax\hbox{$\inbar\kern-.3em{\rm C}$}}
\def\IR{\relax{\rm I\kern-.18em R}}
\def\IP{\relax{\rm I\kern-.18em P}}
\def\IZ{\relax{\rm I\kern-.18em Z}}

%
%%%%%%%%%%%%%%%%%%%%%%%%%%%%%%%%%%%%
%

%
\catcode`\@=11
\def\slash#1{\mathord{\mathpalette\c@ncel{#1}}}\overfullrule=0pt

\def\underrel#1\over#2{\mathrel{\mathop{\kern\z@#1}\limits_{#2}}}

\catcode`\@=12
%%%%%%%%%%%%%%%%%%%%%%%%%%%%%%%%%%%%%%%%%%%%%%%%%%%%%%%%%%%%%%

%

\def \cosh{{\rm cosh}}

\def\exp{{\rm exp}}

%%%%%%%%%%%%%%%%%%%%%%%%%%%%%%%%%%%%%%%%%%%%%%%%%%%%%%%%%%%%%%
% new defs:

%\KutasovER
\lref\KutasovER{
D.~Kutasov and V.~Niarchos,
``Tachyon effective actions in open string theory,''
Nucl.\ Phys.\ B {\bf 666}, 56 (2003)
[arXiv:hep-th/0304045].
%%CITATION = HEP-TH 0304045;%%
}

%\NiarchosRW
\lref\NiarchosRW{
V.~Niarchos,
``Notes on tachyon effective actions and Veneziano amplitudes,''
hep-th/0401066.
%%CITATION = HEP-TH 0401066;%%
}

%\LukyanovNJ
\lref\LukyanovNJ{
  S.~L.~Lukyanov, E.~S.~Vitchev and A.~B.~Zamolodchikov,
  ``Integrable model of boundary interaction: The paperclip,''
  Nucl.\ Phys.\ B {\bf 683}, 423 (2004)
  [arXiv:hep-th/0312168].
  %%CITATION = HEP-TH 0312168;%%
}

%\CallanAT
\lref\CallanAT{
C.~G.~.~Callan, J.~A.~Harvey and A.~Strominger,
``Supersymmetric string solitons,''
arXiv:hep-th/9112030.
%%CITATION = HEP-TH 9112030;%%
}

%\SenAN
\lref\SenAN{
A.~Sen,
``Field theory of tachyon matter,''
Mod.\ Phys.\ Lett.\ A {\bf 17}, 1797 (2002)
[arXiv:hep-th/0204143].
%%CITATION = HEP-TH 0204143;%%
}

%\PolchinskiRQ
\lref\PolchinskiRQ{
J.~Polchinski,
``String Theory. Vol. 1: An Introduction To The Bosonic String,''
Cambridge University Press, 1998.
%\href{http://www.slac.stanford.edu/spires/find/hep/www?irn=4634799}{SPIRES entry}
}

%\PolchinskiRR
\lref\PolchinskiRR{
J.~Polchinski,
``String Theory. Vol. 2: Superstring Theory And Beyond,''
Cambridge University Press, 1998.
%\href{http://www.slac.stanford.edu/spires/find/hep/www?irn=4634802}{SPIRES entry}
}

%\TseytlinDJ
\lref\TseytlinDJ{
A.~A.~Tseytlin,
``Born-Infeld action, supersymmetry and string theory,''
arXiv:hep-th/9908105.
%%CITATION = HEP-TH 9908105;%%
}

%\FelderSV
\lref\FelderSV{
G.~N.~Felder, L.~Kofman and A.~Starobinsky,
``Caustics in tachyon matter and other Born-Infeld scalars,''
JHEP {\bf 0209}, 026 (2002)
[arXiv:hep-th/0208019].
%%CITATION = HEP-TH 0208019;%%
}

%\FelderXU
\lref\FelderXU{
G.~N.~Felder and L.~Kofman,
``Inhomogeneous fragmentation of the rolling tachyon,''
arXiv:hep-th/0403073.
%%CITATION = HEP-TH 0403073;%%
}

%\SenXS
\lref\SenXS{
A.~Sen,
``Open-closed duality at tree level,''
Phys.\ Rev.\ Lett.\  {\bf 91}, 181601 (2003)
[arXiv:hep-th/0306137].
%%CITATION = HEP-TH 0306137;%%
}

%\SenIV
\lref\SenIV{
A.~Sen,
``Open-closed duality: Lessons from matrix model,''
Mod.\ Phys.\ Lett.\ A {\bf 19}, 841 (2004)
[arXiv:hep-th/0308068].
%%CITATION = HEP-TH 0308068;%%
}

%\SenNU
\lref\SenNU{
A.~Sen,
``Rolling tachyon,''
JHEP {\bf 0204}, 048 (2002)
[arXiv:hep-th/0203211].
%%CITATION = HEP-TH 0203211;%%
}

%\SenIN
\lref\SenIN{
A.~Sen,
``Tachyon matter,''
JHEP {\bf 0207}, 065 (2002)
[arXiv:hep-th/0203265].
%%CITATION = HEP-TH 0203265;%%
}

%\LarsenWC
\lref\LarsenWC{
F.~Larsen, A.~Naqvi and S.~Terashima,
``Rolling tachyons and decaying branes,''
JHEP {\bf 0302}, 039 (2003)
[arXiv:hep-th/0212248].
%%CITATION = HEP-TH 0212248;%%
}

%\OkudaYD
\lref\OkudaYD{
T.~Okuda and S.~Sugimoto,
``Coupling of rolling tachyon to closed strings,''
Nucl.\ Phys.\ B {\bf 647}, 101 (2002)
[arXiv:hep-th/0208196].
%%CITATION = HEP-TH 0208196;%%
}

%\LambertZR
\lref\LambertZR{
N.~Lambert, H.~Liu and J.~Maldacena,
``Closed strings from decaying D-branes,''
arXiv:hep-th/0303139.
%%CITATION = HEP-TH 0303139;%%
}

%\GaiottoRM
\lref\GaiottoRM{
D.~Gaiotto, N.~Itzhaki and L.~Rastelli,
``Closed strings as imaginary D-branes,''
arXiv:hep-th/0304192.
%%CITATION = HEP-TH 0304192;%%
}

%\SenMD
\lref\SenMD{
A.~Sen,
``Supersymmetric world-volume action for non-BPS D-branes,''
JHEP {\bf 9910}, 008 (1999)
[arXiv:hep-th/9909062].
%%CITATION = HEP-TH 9909062;%%
}

%\GarousiTR
\lref\GarousiTR{
M.~R.~Garousi,
``Tachyon couplings on non-BPS D-branes and Dirac-Born-Infeld action,''
Nucl.\ Phys.\ B {\bf 584}, 284 (2000)
[arXiv:hep-th/0003122].
%%CITATION = HEP-TH 0003122;%%
}

%\BergshoeffDQ
\lref\BergshoeffDQ{
E.~A.~Bergshoeff, M.~de Roo, T.~C.~de Wit, E.~Eyras and S.~Panda,
``T-duality and actions for non-BPS D-branes,''
JHEP {\bf 0005}, 009 (2000)
[arXiv:hep-th/0003221].
%%CITATION = HEP-TH 0003221;%%
}

%\KlusonIY
\lref\KlusonIY{
J.~Kluson,
``Proposal for non-BPS D-brane action,''
Phys.\ Rev.\ D {\bf 62}, 126003 (2000)
[arXiv:hep-th/0004106].
%%CITATION = HEP-TH 0004106;%%
}

%\SenTM
\lref\SenTM{
A.~Sen,
``Dirac-Born-Infeld action on the tachyon kink and vortex,''
Phys.\ Rev.\ D {\bf 68}, 066008 (2003)
[arXiv:hep-th/0303057].
%%CITATION = HEP-TH 0303057;%%
}

%\SenMV
\lref\SenMV{
A.~Sen,
``Remarks on tachyon driven cosmology,''
arXiv:hep-th/0312153.
%%CITATION = HEP-TH 0312153;%%
}

%\ElitzurPQ
\lref\ElitzurPQ{
S.~Elitzur, A.~, D.~Kutasov, E.~Rabinovici and G.~Sarkissian,
``D-branes in the background of NS fivebranes,''
JHEP {\bf 0008}, 046 (2000)
[arXiv:hep-th/0005052].
%%CITATION = HEP-TH 0005052;%%
}

%\PelcKB
\lref\PelcKB{
O.~Pelc,
``On the quantization constraints for a D3 brane in the geometry of NS5
%branes,''
JHEP {\bf 0008}, 030 (2000)
[arXiv:hep-th/0007100].
%%CITATION = HEP-TH 0007100;%%
}

%\RibaultSG
\lref\RibaultSG{
S.~Ribault,
``D3-branes in NS5-branes backgrounds,''
hep-th/0301092,
JHEP {\bf 0302}, 044 (2003)
.
%%CITATION = HEP-TH 0301092;%%
}

%\SeibergZK
\lref\SeibergZK{
N.~Seiberg,
``New theories in six dimensions and matrix description of M-theory on  T**5
%and T**5/Z(2),''
Phys.\ Lett.\ B {\bf 408}, 98 (1997)
[arXiv:hep-th/9705221].
%%CITATION = HEP-TH 9705221;%%
}

%\GutperleAI
\lref\GutperleAI{
M.~Gutperle and A.~Strominger,
``Spacelike branes,''
JHEP {\bf 0204}, 018 (2002)
[arXiv:hep-th/0202210].
%%CITATION = HEP-TH 0202210;%%
}

%\GibbonsMD
\lref\GibbonsMD{
G.~W.~Gibbons,
``Cosmological evolution of the rolling tachyon,''
Phys.\ Lett.\ B {\bf 537}, 1 (2002)
[arXiv:hep-th/0204008].
%%CITATION = HEP-TH 0204008;%%
}

%\ShiuXP
\lref\ShiuXP{
G.~Shiu, S.~H.~H.~Tye and I.~Wasserman,
``Rolling tachyon in brane world cosmology from superstring field theory,''
Phys.\ Rev.\ D {\bf 67}, 083517 (2003)
[arXiv:hep-th/0207119].
%%CITATION = HEP-TH 0207119;%%
}

%\GiveonZM
\lref\GiveonZM{
A.~Giveon, D.~Kutasov and O.~Pelc,
``Holography for non-critical superstrings,''
JHEP {\bf 9910}, 035 (1999)
[arXiv:hep-th/9907178].
%%CITATION = HEP-TH 9907178;%%
}

%\AharonyUB
\lref\AharonyUB{
O.~Aharony, M.~Berkooz, D.~Kutasov and N.~Seiberg,
``Linear dilatons, NS5-branes and holography,''
JHEP {\bf 9810}, 004 (1998)
[arXiv:hep-th/9808149].
%%CITATION = HEP-TH 9808149;%%
}

%\GiveonPX
\lref\GiveonPX{
A.~Giveon and D.~Kutasov,
``Little string theory in a double scaling limit,''
JHEP {\bf 9910}, 034 (1999)
[arXiv:hep-th/9909110].
%%CITATION = HEP-TH 9909110;%%
}

%\SenMG
\lref\SenMG{
A.~Sen,
``Non-BPS states and branes in string theory,''
arXiv:hep-th/9904207.
%%CITATION = HEP-TH 9904207;%%
}

%\KutasovDJ
\lref\KutasovDJ{
D.~Kutasov,
``D-brane dynamics near NS5-branes,''
arXiv:hep-th/0405058.
%%CITATION = HEP-TH 0405058;%%
}

%\MinahanTG
\lref\MinahanTG{
J.~A.~Minahan and B.~Zwiebach,
``Gauge fields and fermions in tachyon effective field theories,''
JHEP {\bf 0102}, 034 (2001)
[arXiv:hep-th/0011226].
%%CITATION = HEP-TH 0011226;%%
}

%\SenNF
\lref\SenNF{
 A.~Sen,
 ``Tachyon dynamics in open string theory,''
  arXiv:hep-th/0410103.
  %%CITATION = HEP-TH 0410103;%%
}

%\KutasovDJ
\lref\KutasovDJ{
  D.~Kutasov,
  ``D-brane dynamics near NS5-branes,''
  arXiv:hep-th/0405058.
  %%CITATION = HEP-TH 0405058;%%
}

%\NakayamaYX
\lref\NakayamaYX{
  Y.~Nakayama, Y.~Sugawara and H.~Takayanagi,
  ``Boundary states for the rolling D-branes in NS5 background,''
  JHEP {\bf 0407}, 020 (2004)
  [arXiv:hep-th/0406173].
  %%CITATION = HEP-TH 0406173;%%
}

%\SahakyanCQ
\lref\SahakyanCQ{
  D.~A.~Sahakyan,
  ``Comments on D-brane dynamics near NS5-branes,''
  JHEP {\bf 0410}, 008 (2004)
  [arXiv:hep-th/0408070].
  %%CITATION = HEP-TH 0408070;%%
}

%\NakayamaGE
\lref\NakayamaGE{
  Y.~Nakayama, K.~L.~Panigrahi, S.~J.~Rey and H.~Takayanagi,
  ``Rolling down the throat in NS5-brane background: The case of electrified
  D-brane,''
  JHEP {\bf 0501}, 052 (2005)
  [arXiv:hep-th/0412038].
  %%CITATION = HEP-TH 0412038;%%
}

%\YavartanooWB
\lref\YavartanooWB{
  H.~Yavartanoo,
  ``Cosmological solution from D-brane motion in NS5-branes background,''
  arXiv:hep-th/0407079.
  %%CITATION = HEP-TH 0407079;%%
}

%\PanigrahiQR
\lref\PanigrahiQR{
  K.~L.~Panigrahi,
  ``D-brane dynamics in Dp-brane background,''
  Phys.\ Lett.\ B {\bf 601}, 64 (2004)
  [arXiv:hep-th/0407134].
  %%CITATION = HEP-TH 0407134;%%
}

%\GhodsiWN
\lref\GhodsiWN{
  A.~Ghodsi and A.~E.~Mosaffa,
  ``D-brane dynamics in RR deformation of NS5-branes background and tachyon
  cosmology,''
  Nucl.\ Phys.\ B {\bf 714}, 30 (2005)
  [arXiv:hep-th/0408015].
  %%CITATION = HEP-TH 0408015;%%
}

%\KutasovCT
\lref\KutasovCT{
  D.~Kutasov,
  ``A geometric interpretation of the open string tachyon,''
  arXiv:hep-th/0408073.
  %%CITATION = HEP-TH 0408073;%%
}

%\YogendranDM
\lref\YogendranDM{
  K.~P.~Yogendran,
  ``D-branes in 2D Lorentzian black hole,''
  JHEP {\bf 0501}, 036 (2005)
  [arXiv:hep-th/0408114].
  %%CITATION = HEP-TH 0408114;%%
}

%\SaremiYD
\lref\SaremiYD{
  O.~Saremi, L.~Kofman and A.~W.~Peet,
  ``Folding branes,''
  Phys.\ Rev.\ D {\bf 71}, 126004 (2005)
  [arXiv:hep-th/0409092].
  %%CITATION = HEP-TH 0409092;%%
}

%\KlusonXC
\lref\KlusonXC{
  J.~Kluson,
  ``Non-BPS D-brane near NS5-branes,''
  JHEP {\bf 0411}, 013 (2004)
  [arXiv:hep-th/0409298].
  %%CITATION = HEP-TH 0409298;%%
}

%\ToumbasFE
\lref\ToumbasFE{
  N.~Toumbas and J.~Troost,
  ``A time-dependent brane in a cosmological background,''
  JHEP {\bf 0411}, 032 (2004)
  [arXiv:hep-th/0410007].
  %%CITATION = HEP-TH 0410007;%%
}

%\KlusonYK
\lref\KlusonYK{
  J.~Kluson,
  ``Non-BPS Dp-brane in the background of NS5-branes on transverse R**3 x
  S**1,''
  JHEP {\bf 0503}, 032 (2005)
  [arXiv:hep-th/0411014].
  %%CITATION = HEP-TH 0411014;%%
}

%\BakTP
\lref\BakTP{
  D.~Bak, S.~J.~Rey and H.~U.~Yee,
  ``Exactly soluble dynamics of (p,q) string near macroscopic fundamental
  strings,''
  JHEP {\bf 0412}, 008 (2004)
  [arXiv:hep-th/0411099].
  %%CITATION = HEP-TH 0411099;%%
}

%\ThomasCD
\lref\ThomasCD{
  S.~Thomas and J.~Ward,
  ``D-brane dynamics and NS5 rings,''
  JHEP {\bf 0502}, 015 (2005)
  [arXiv:hep-th/0411130].
  %%CITATION = HEP-TH 0411130;%%
}

%\ChenVW
\lref\ChenVW{
  B.~Chen, M.~Li and B.~Sun,
  ``Dbrane near NS5-branes: With electromagnetic field,''
  JHEP {\bf 0412}, 057 (2004)
  [arXiv:hep-th/0412022].
  %%CITATION = HEP-TH 0412022;%%
}

%\KlusonQX
\lref\KlusonQX{
  J.~Kluson,
  ``Non-BPS Dp-brane in Dk-brane background,''
  JHEP {\bf 0503}, 044 (2005)
  [arXiv:hep-th/0501010].
  %%CITATION = HEP-TH 0501010;%%
}

%\LapanQZ
\lref\LapanQZ{
  J.~M.~Lapan and W.~Li,
  ``Falling D0-branes in 2D superstring theory,''
  arXiv:hep-th/0501054.
  %%CITATION = HEP-TH 0501054;%%
}

%\ChenWM
\lref\ChenWM{
  B.~Chen and B.~Sun,
  ``Note on DBI dynamics of Dbrane near NS5-branes,''
  arXiv:hep-th/0501176.
  %%CITATION = HEP-TH 0501176;%%
}

%\ThomasAM
\lref\ThomasAM{
  S.~Thomas and J.~Ward,
  ``D-brane dynamics near compactified NS5-branes,''
  JHEP {\bf 0506}, 062 (2005)
  [arXiv:hep-th/0501192].
  %%CITATION = HEP-TH 0501192;%%
}

%\HuangRD
\lref\HuangRD{
  W.~H.~Huang,
  ``Tubular solutions in NS5-brane, Dp-brane and macroscopic strings
  background,''
  JHEP {\bf 0502}, 061 (2005)
  [arXiv:hep-th/0502023].
  %%CITATION = HEP-TH 0502023;%%
}

%\KlusonJR
\lref\KlusonJR{
  J.~Kluson,
  ``Note about non-BPS and BPS Dp-branes in near horizon region of N
  Dk-branes,''
  JHEP {\bf 0503}, 071 (2005)
  [arXiv:hep-th/0502079].
  %%CITATION = HEP-TH 0502079;%%
}

%\ThomasFW
\lref\ThomasFW{
S.~Thomas and J.~Ward,
``Geometrical tachyon kinks and NS5 branes,''
arXiv:hep-th/0502228.
%%CITATION = HEP-TH 0502228;%%
}

%\ThomasFU
\lref\ThomasFU{
S.~Thomas and J.~Ward,
``Inflation from geometrical tachyons,''
arXiv:hep-th/0504226.
%%CITATION = HEP-TH 0504226;%%
}

%\KlusonAB
\lref\KlusonAB{
J.~Kluson and K.~L.~Panigrahi,
``Supertube dynamics in diverse backgrounds,''
arXiv:hep-th/0506012.
%%CITATION = HEP-TH 0506012;%%
}

%\NakayamaPK
\lref\NakayamaPK{
Y.~Nakayama, S.~J.~Rey and Y.~Sugawara,
``D-brane propagation in two-dimensional black hole geometries,''
arXiv:hep-th/0507040.
%%CITATION = HEP-TH 0507040;%%
}

%\FateevTK
\lref\FateevTK{
V.~A.~Fateev, E.~Onofri and A.~B.~Zamolodchikov,
``The Sausage model (integrable deformations of O(3) sigma model),''
Nucl.\ Phys.\ B {\bf 406}, 521 (1993).
%%CITATION = NUPHA,B406,521;%%
}

\lref\fzz{V.~A.~Fateev, A.~B.~Zamolodchikov and Al.~B.~Zamolodchikov, unpublished.}

%\KazakovPM
\lref\KazakovPM{  
V.~Kazakov, I.~K.~Kostov and D.~Kutasov,  
``A matrix model for the two-dimensional black hole,''  
Nucl.\ Phys.\ B {\bf 622}, 141 (2002)
[arXiv:hep-th/0101011].
%%CITATION = HEP-TH 0101011;%%
}

%\GiveonUP
\lref\GiveonUP{
A.~Giveon and D.~Kutasov,
``Notes on AdS(3),''
Nucl.\ Phys.\ B {\bf 621}, 303 (2002)
[arXiv:hep-th/0106004].
%%CITATION = HEP-TH 0106004;%%
}

%\HoriAX
\lref\HoriAX{
K.~Hori and A.~Kapustin,
``Duality of the fermionic 2d black hole and N = 2 Liouville theory as
mirror symmetry,''
JHEP {\bf 0108}, 045 (2001)
[arXiv:hep-th/0104202].
%%CITATION = HEP-TH 0104202;%%
}

%\AharonyVK
\lref\AharonyVK{
O.~Aharony, B.~Fiol, D.~Kutasov and D.~A.~Sahakyan,
``Little string theory and heterotic/type II duality,''
Nucl.\ Phys.\ B {\bf 679}, 3 (2004)
[arXiv:hep-th/0310197].
%%CITATION = HEP-TH 0310197;%%
}

%\AharonyXN
\lref\AharonyXN{
O.~Aharony, A.~Giveon and D.~Kutasov,
``LSZ in LST,''
Nucl.\ Phys.\ B {\bf 691}, 3 (2004)
[arXiv:hep-th/0404016].
%%CITATION = HEP-TH 0404016;%%
}

\lref\SusskindWS{
L.~Susskind,
``Strings, black holes and Lorentz contraction,''
Phys.\ Rev.\ D {\bf 49}, 6606 (1994)
[arXiv:hep-th/9308139];
``Some speculations about black hole entropy in string theory,''
arXiv:hep-th/9309145.
%%CITATION = HEP-TH 9309145;%%
}

%\HorowitzNW
\lref\HorowitzNW{
G.~T.~Horowitz and J.~Polchinski,
``A correspondence principle for black holes and strings,''
Phys.\ Rev.\ D {\bf 55}, 6189 (1997)
[arXiv:hep-th/9612146].
%%CITATION = HEP-TH 9612146;%%
}

%\DijkgraafBA
\lref\DijkgraafBA{
R.~Dijkgraaf, H.~L.~Verlinde and E.~P.~Verlinde,
``String propagation in a black hole geometry,''
Nucl.\ Phys.\ B {\bf 371}, 269 (1992).
%%CITATION = NUPHA,B371,269;%%
}

%\BarsSR
\lref\BarsSR{
I.~Bars and K.~Sfetsos,
``Conformally exact metric and dilaton in string theory on curved
space-time,''
Phys.\ Rev.\ D {\bf 46}, 4510 (1992)
[arXiv:hep-th/9206006].
%%CITATION = HEP-TH 9206006;%%
}

%\TseytlinMY
\lref\TseytlinMY{
A.~A.~Tseytlin,
``Conformal sigma models corresponding to gauged Wess-Zumino-Witten
theories,''
Nucl.\ Phys.\ B {\bf 411}, 509 (1994)
[arXiv:hep-th/9302083].
%%CITATION = HEP-TH 9302083;%%
}
	  
%\UnruhDB
\lref\UnruhDB{
  W.~G.~Unruh,
  ``Notes On Black Hole Evaporation,''
  Phys.\ Rev.\ D {\bf 14}, 870 (1976).
  %%CITATION = PHRVA,D14,870;%%
}

%\McGuiganTG
\lref\McGuiganTG{
  M.~McGuigan,
  ``Finite black hole entropy and string theory,''
  Phys.\ Rev.\ D {\bf 50}, 5225 (1994)
  [arXiv:hep-th/9406201].
  %%CITATION = HEP-TH 9406201;%%
}

%\UnruhIC
\lref\UnruhIC{
  W.~G.~Unruh and R.~M.~Wald,
  ``Acceleration Radiation And Generalized Second Law Of Thermodynamics,''
  Phys.\ Rev.\ D {\bf 25}, 942 (1982).
  %%CITATION = PHRVA,D25,942;%%
}

%\GiveonMI
\lref\GiveonMI{
  A.~Giveon, D.~Kutasov, E.~Rabinovici and A.~Sever,
  ``Phases of quantum gravity in AdS(3) and linear dilaton backgrounds,''
  Nucl.\ Phys.\ B {\bf 719}, 3 (2005)
  [arXiv:hep-th/0503121].
  %%CITATION = HEP-TH 0503121;%%
}

%\ElitzurCB
\lref\ElitzurCB{
S.~Elitzur, A.~Forge and E.~Rabinovici,
``Some global aspects of string compactifications,''
Nucl.\ Phys.\ B {\bf 359}, 581 (1991).
%%CITATION = NUPHA,B359,581;%%
}

%\WittenYR
\lref\WittenYR{
E.~Witten,
``On string theory and black holes,''
Phys.\ Rev.\ D {\bf 44}, 314 (1991).
%%CITATION = PHRVA,D44,314;%%
}

%\MandalTZ
\lref\MandalTZ{
G.~Mandal, A.~M.~Sengupta and S.~R.~Wadia,
``Classical solutions of two-dimensional string theory,''
Mod.\ Phys.\ Lett.\ A {\bf 6}, 1685 (1991).
%%CITATION = MPLAE,A6,1685;%%
}

%\GiveonTQ
\lref\GiveonTQ{
A.~Giveon and D.~Kutasov,
``Comments on double scaled little string theory,''
JHEP {\bf 0001}, 023 (2000)
[arXiv:hep-th/9911039].
%%CITATION = HEP-TH 9911039;%%
}

%\HorowitzJC
\lref\HorowitzJC{
G.~T.~Horowitz and J.~Polchinski,
``Self gravitating fundamental strings,''
Phys.\ Rev.\ D {\bf 57}, 2557 (1998)
[arXiv:hep-th/9707170].
%%CITATION = HEP-TH 9707170;%%
}

%\DamourAW
\lref\DamourAW{
T.~Damour and G.~Veneziano,
``Self-gravitating fundamental strings and black holes,''
Nucl.\ Phys.\ B {\bf 568}, 93 (2000)
[arXiv:hep-th/9907030].
%%CITATION = HEP-TH 9907030;%%
}

%\FelderXU
\lref\FelderXU{
G.~N.~Felder and L.~Kofman,
``Inhomogeneous fragmentation of the rolling tachyon,''
Phys.\ Rev.\ D {\bf 70}, 046004 (2004)
[arXiv:hep-th/0403073].
%%CITATION = HEP-TH 0403073;%%
}

%\SaremiYD
\lref\SaremiYD{
O.~Saremi, L.~Kofman and A.~W.~Peet,
``Folding branes,''
Phys.\ Rev.\ D {\bf 71}, 126004 (2005)
[arXiv:hep-th/0409092].
%%CITATION = HEP-TH 0409092;%%
}

%\AtickSI
\lref\AtickSI{
J.~J.~Atick and E.~Witten,
``The Hagedorn Transition And The Number Of Degrees Of Freedom Of String
Theory,''
 Nucl.\ Phys.\ B {\bf 310}, 291 (1988).
 %%CITATION = NUPHA,B310,291;%%
}

%\BarbonDD
\lref\BarbonDD{
J.~L.~F.~Barbon and E.~Rabinovici,
``Touring the Hagedorn ridge,''
arXiv:hep-th/0407236.
%%CITATION = HEP-TH 0407236;%%
}

%\rightline{EFI-04-..}
\Title{
}
{\vbox{\centerline{Accelerating Branes and the String/Black Hole Transition}}}
\bigskip
\centerline{David Kutasov}
\bigskip
\centerline{{\it Enrico Fermi Inst. and Dept. of Physics,
University of Chicago}}
\centerline{\it 5640 S. Ellis Ave., Chicago, IL 60637-1433, USA}
\bigskip\bigskip\bigskip
\noindent
String theory in Euclidean flat space with a spacelike linear dilaton
contains a $D1$-brane which looks like a semi-infinite hairpin. In 
addition to its curved shape, this ``hairpin brane'' has a condensate 
of the open string tachyon stretched between its two sides. The tachyon 
smears the brane and shifts the location of its tip. 
The Minkowski continuation of the hairpin brane describes 
a $D0$-brane freely falling in a linear dilaton background. Effects that 
in Euclidean space are attributed to the tachyon condensate, give rise in 
the Minkowski case to a stringy smearing of the trajectory of the $D$-brane 
by an amount that grows as its acceleration increases. When the Unruh 
temperature of the brane reaches the Hagedorn temperature of perturbative 
string theory in the throat, the rolling $D$-brane state becomes 
non-normalizable. We propose that black holes in string theory exhibit 
similar properties. The Euclidean black hole solution has a condensate 
of a tachyon winding around Euclidean time. The Minkowski manifestation of 
this condensate is a smearing of the geometry in a layer around the horizon. 
As the Hawking temperature, $T_{bh}$, increases, the width of this layer grows. 
When $T_{bh}$ reaches the Hagedorn temperature, the size of this ``smeared 
horizon'' diverges, and the black hole becomes non-normalizable. This provides 
a new point of view on the string/black hole transition.

\vfill

\Date{}

%%%%%%%%%%%%%%%%%%%%%%%%%%%%%%%%%%%%%%%%%%%%%%%%%%%%%%%%%%%%%%%%%%%%%%
%%%%%%%%%%%%%%%%%%%%%%%%%%%%%%%%%%%%%%%%%%%%%%%%%%%%%%%%%%%%%%%%%%%%%%
\newsec{Introduction}

Linear dilaton backgrounds arise naturally in string theory
as near-horizon (or throat) geometries of Neveu-Schwarz 
fivebranes and other singularities \refs{\CallanAT,\GiveonZM}. 
In the last year there has been some work on the real time 
dynamics of $D$-branes in such backgrounds. The fact that 
the string coupling increases as one approaches the singularity
leads to an attractive potential for any $D$-brane localized in the
throat, which causes it to roll into the strong coupling region. 

The resulting trajectory was analyzed in \KutasovDJ\ by using the 
Dirac-Born-Infeld (DBI) action for the $D$-brane. It was also pointed 
out in \KutasovDJ\ that the exact boundary state describing the rolling 
D-brane is a Wick rotated version of the ``hairpin brane'' constructed 
previously in \LukyanovNJ, or  of its generalization to the superstring. 
This generalization was subsequently studied in 
\refs{\NakayamaYX\SahakyanCQ-\NakayamaGE}.
Some additional interesting related work appeared in 
\refs{\YavartanooWB\PanigrahiQR\GhodsiWN\KutasovCT\YogendranDM
\SaremiYD\KlusonXC\ToumbasFE\KlusonYK\BakTP\ThomasCD\ChenVW
\KlusonQX\LapanQZ\ChenWM\ThomasAM\HuangRD\KlusonJR\ThomasFW
\ThomasFU\KlusonAB-\NakayamaPK}. 

As discussed in \LukyanovNJ, the hairpin brane is an open string analog of the 
cigar CFT,  $SL(2,\IR)/U(1)$. More generally, the ``paperclip'' model \LukyanovNJ, 
whose UV limit is the hairpin brane, is a boundary analog  of the ``sausage'' model  
\FateevTK, whose UV limit is the cigar.  The bulk counterpart of the Wick rotation 
that takes the hairpin to the rolling D-brane, takes the cigar (or Euclidean two 
dimensional black hole) to the Minkowski two dimensional black hole. Thus, some 
aspects of the dynamics of the black hole have open string counterparts. One of 
the motivations for the present work is to use the rolling D-brane system to learn 
about black holes.

For the bulk case, it is known that while for small dilaton slope the Euclidean
$SL(2,\IR)/U(1)$ conformal field theory is well described by a sigma model on 
the cigar, for large slope a better description is in terms of Sine-Liouville CFT 
in the bosonic case \refs{\fzz,\KazakovPM}, or $N=2$ Liouville theory in the 
worldsheet supersymmetric one \GiveonPX. In this description, the cigar is replaced 
by an infinite cylinder, and strings are prevented from exploring the strong coupling 
region by a condensate of a closed string tachyon which winds around the compact 
cycle of the cylinder and increases towards the strong coupling region. 

The cigar-Liouville duality is a strong -- weak coupling duality on the worldsheet. For 
small dilaton gradient, the sigma model on the cigar is weakly coupled since the curvature 
and dilaton gradient are small. For large gradient it is strongly coupled, but in that regime 
the Liouville description is weakly coupled, since the Liouville potential is slowly 
varying. In general, one has to include both the curved metric and the condensate 
of the winding tachyon to get a full description of the physics. Evidence for this
picture comes from both worldsheet \refs{\HoriAX,\GiveonUP} and spacetime Little String 
Theory \refs{\AharonyVK,\AharonyXN} considerations. 

The close relation between the cigar and its boundary counterpart, the hairpin,
suggests \LukyanovNJ\ that a similar strong-weak coupling duality exists for
the hairpin as well. In this case the statement is that for small dilaton gradient
the boundary state corresponds to a brane whose shape is described by the 
hairpin, while for large gradient it corresponds to a brane and anti-brane at a 
small distance determined by the gradient, with a condensate of the open string 
tachyon stretched between them. We will describe this duality in more detail below. 

The fact that the cigar and hairpin CFT's contain a condensate of a winding tachyon
leads to an interesting question regarding the  Minkowski continuations of these solutions. 
From the point of view of these continuations, the tachyon has non-zero winding
in the Euclidean time direction, and it is not clear how to interpret it after continuation
to Minkowski space. This is especially puzzling since in Euclidean space the tachyon
condensate has a large effect when the dilaton gradient is of order one. In particular,
above a critical value of the gradient, the hairpin and cigar states become non-normalizable
as a consequence of its existence. 

The purpose of this paper is to clarify this issue. Our discussion has two parts. In the
first (sections 2, 3) we consider the hairpin brane and its Minkowski continuation. We 
identify the effects due to the winding tachyon condensate in both Euclidean and 
Minkowski space, and study these effects as a function of the gradient of the dilaton. 
In Euclidean space, the tachyon smears the hairpin shape by an amount whose size is
determined by the region in which the mass squared of the tachyon is negative. The size of 
this region grows with the gradient of the dilaton, and diverges when the asymptotic mass 
of the tachyon goes to zero. At that point, the brane ceases to be normalizable.

The Minkowski continuation of the hairpin brane corresponds to a $D$-brane
freely falling in the linear dilaton background. The smearing of the hairpin 
brane that is due to the condensate of the stretched tachyon gives rise 
after Wick rotation to a stringy smearing of the trajectory of the rolling 
$D$-brane. The size of this smearing grows with the acceleration of the 
$D$-brane. It diverges when the Unruh temperature of the brane approaches the 
perturbative Hagedorn temperature in the linear dilaton throat. At that point 
the rolling $D$-brane boundary state is pushed out to $\phi\to\infty$ and ceases 
to be normalizable.  

We propose the following qualitative picture to explain this smearing. In 
perturbative string theory, the excitations of the $D$-brane are described 
by open strings both of whose ends lie on the $D$-brane. When the brane is 
at rest or moving at a constant velocity, its ground state has no open strings 
excited, and the location of the brane can be sharply defined. 
In our case, the continuation from Euclidean space describes a state of the 
accelerating brane in which there are open strings which oscillate along the 
$\phi$ axis, providing a kind of stringy halo to the brane.

The acceleration in the negative $\phi$ direction gives rise to a force on 
these strings in the opposite, positive $\phi$, direction. This shifts the halo, 
on average, to a larger value of $\phi$ than the original $D$-brane. The size 
of this halo increases with the acceleration. When the Unruh temperature of the
brane reaches the Hagedorn temperature, the halo extends all the way to 
$\phi\to\infty$ and pushes the brane out of the linear dilaton throat.  

The second part of our discussion (section 4) concerns black holes. For the two 
dimensional $SL(2,\IR)/U(1)$ black hole we argue that in Euclidean space the 
winding tachyon gives rise to stringy corrections which smear the geometry 
of the cigar. The smearing is again determined by the region in which 
the mass squared of the wrapped tachyon is negative. 

After Wick rotation, this results in a smearing of the region near the horizon of
the Minkowski black hole. The smeared region is characterized by the fact that the
local Hawking temperature in it exceeds the Hagedorn temperature. For small dilaton
gradient, the size of this region is of order one in string units. One can think of 
it as a kind of stringy stretched horizon \SusskindWS. As the dilaton gradient 
increases, the size of the smeared region grows. When the Hawking temperature at 
infinity approaches the perturbative Hagedorn temperature in the throat, this size 
diverges and the black hole becomes non-normalizable. 

Like in the case of the accelerating $D$-brane, we argue that the reason for the
appearance of the stretched horizon in the Lorentzian black hole solution is a
condensate of closed strings near the horizon. These strings are the Lorentzian 
manifestation of the winding tachyon condensate. When the Hawking temperature 
is close to the Hagedorn temperature, the winding tachyon is light everywhere, and in 
Minkowski space the closed strings that form the stretched horizon are highly excited. 

We propose that the winding tachyon condensate is a general property
of classical black hole solutions in string theory. For example, in the $d$ 
dimensional Euclidean Schwarzschild solution, the asymptotic radius of Euclidean 
time (or inverse Hawking temperature) grows with the mass of the black hole $M$. 
The boundary conditions are such that a tachyon winding once around Euclidean 
time survives the GSO projection, but its mass, $m_\infty(M)$ is large in this 
limit. Therefore, the tachyon condensate goes rapidly to zero at large distances,  
$T(r)\sim\exp(-m_\infty r)$. However, in a small layer near the horizon, which is 
again characterized by the fact that in it the local Hawking temperature exceeds 
the Hagedorn temperature, the tachyon condensate is appreciable, and provides some 
stringy smearing of the solution, as in \SusskindWS. As the mass of the black hole 
decreases, the width of this stretched horizon grows until, when the Hawking 
temperature of the black hole approaches the Hagedorn temperature, it extends all 
the way to infinity, and the black hole becomes non-normalizable.

The above discussion also sheds light on the string/black hole correspondence
of \refs{\SusskindWS,\HorowitzNW}. The conventional picture is that the transition
from black holes to strings happens when the curvature near the horizon becomes
of order the string scale. A more precise formulation suggested by our considerations 
is that the transition takes place when the Hawking temperature of the black hole 
becomes equal to the perturbative Hagedorn temperature of fundamental strings in 
the spacetime without the black hole. 

Near the transition, the black hole typically has horizon size of order $l_s$, 
but its stretched horizon has a much larger size, and as the Hawking temperature 
approaches the Hagedorn temperature, it looks more and more like a highly excited 
string state. This explains why, at the transition, its entropy must agree with 
that of a perturbative string state with the same mass. In order to find the value 
of the mass at which the transition takes place, one needs to know the exact 
relation between the mass and the Hawking temperature in this regime, but if one 
can construct an exact classical black hole solution in string theory, this
formulation gives precise predictions that can in principle be verified.

\newsec{Hairpin}

In this section we describe some features of the hairpin boundary state,
and its dual description in terms of boundary $N=2$ Liouville. We restrict the 
discussion to the worldsheet supersymmetric case (the boundary analog of 
\GiveonPX) since this is the case relevant for the superstring. We will not discuss
the bosonic version of this construction (the boundary analog of \refs{\fzz,\KazakovPM}).

We start with a flat Euclidean two dimensional space labeled by $(\phi,x)$,
\eqn\rphirx{\IR_\phi\times \IR_x~.}
The dilaton depends linearly on $\phi$,
\eqn\lindil{\Phi=-{Q\over2}\phi~.}
We set $\alpha'=2$, such that the central charge of $\phi$ is
$c_\phi=1+3Q^2$. Since we are planning to embed \rphirx\ in the superstring,
there are also two free fermions, $(\psi_\phi,\psi_x)$ and their counterparts 
with the other worldsheet chirality. 

The shape of the hairpin brane is described by the relation
\eqn\hairp{e^{-{Q\over2}\phi}=2C\cos{Q\over2}x~.}
The constant $C$ can be set to one by shifting $\phi$, and we will often 
do so below. As explained in \refs{\KutasovDJ,\LukyanovNJ}, \hairp\ is 
obtained by solving the equations of motion of the DBI action for a 
$D1$-brane in the linear dilaton background \rphirx. This action follows 
from the one loop $\beta$-function of the boundary RG. Apriori, one might 
expect that it is accurate only for small $Q$ and receives $\alpha'$ 
corrections. For the bosonic analog of this construction this is indeed the 
case \LukyanovNJ, but for the fermionic one \hairp\ is expected to be exact, 
in a sense that will be made more precise below. 

One of the reasons for this expectation is the analogy with the bulk problem. As 
mentioned in the introduction, \hairp\ can be thought of as an open string analog 
of the Euclidean cigar CFT, $SL(2,\IR)/U(1)$. In that case, it is known that while in 
the bosonic string the geometry receives $\alpha'$ corrections \DijkgraafBA, in the 
fermionic string it is one loop exact \refs{\BarsSR,\TseytlinMY} since the background 
preserves worldsheet $(2,2)$ superconformal symmetry. The hairpin brane also 
preserves $N=2$ superconformal symmetry on the worldsheet and is closely related 
to the cigar. It is natural to expect that in this case too, \hairp\ is exact. 

The hairpin shape \hairp\ has the following properties. As $\phi\to\infty$ (\ie\
as we approach the boundary of the linear dilaton space \rphirx), 
$x\to\pm {\pi\over Q}$. In this region the system looks like a $D1$-brane and an anti 
$D1$-brane separated by the distance $\delta x={2\pi\over Q}$. As we move to 
smaller $\phi$, the two $D$-branes curve towards each other as follows:
\eqn\deltax{\delta x={2\pi\over Q}-\lambda e^{-{Q\over2}\phi}+O(e^{-Q\phi})~,}
where $\lambda$ is a positive constant. They meet at $x=0$, where the string
coupling takes its largest value along the brane. Since we can choose this value
to be arbitrarily small, we expect the boundary CFT describing the hairpin brane
to be well behaved on the disk. 

An important probe of the hairpin brane is the disk one point function of 
the bulk operators
\eqn\bulkt{T(p,q)=e^{(-{Q\over2}+ip)\phi+iqx}~.} 
These operators have worldsheet scaling dimension 
$\Delta_{p,q}=\bar\Delta_{p,q}={1\over2}p^2+{1\over2}q^2+{1\over8}Q^2$.
Their one point function is given by \refs{\LukyanovNJ,\NakayamaYX}
\eqn\onept{\Psi(p,q)=\langle T(p,q)\rangle\simeq {\Gamma(1-iQp)\Gamma(-i{2p\over Q})
\over
\Gamma(\half-i{p\over Q}+{q\over Q})\Gamma(\half-i{p\over Q}-{q\over Q})}}
where we neglected an overall constant that will not play a role below. 

The one point function \onept\ determines the coefficient of the identity in the 
hairpin boundary state:
\eqn\hairbound{|{\rm hairpin}\rangle=\Psi(\phi,x)|0\rangle+\cdots}
where the $\cdots$ stands for oscillator states, and
\eqn\wavefun{\Psi(\phi,x)= \int {dpdq\over (2\pi)^2}e^{-ip\phi-iqx}\Psi(p,q)}
is the position space wave function of the hairpin.

In the limit $Q\to 0$, the wave function $\Psi(p,q)$ \onept\ can be approximated 
by  
\eqn\psicl{
\Psi_{\rm cl}(p,q)\simeq{\Gamma(-i{2p\over Q})\over
\Gamma(\half-i{p\over Q}+{q\over Q})\Gamma(\half-i{p\over Q}-{q\over Q})}~.}
Using this classical wave function in \wavefun, gives \NakayamaYX
\eqn\classwave{
\Psi_{\rm cl}(\phi,x)=\int {dpdq\over (2\pi)^2}e^{-ip\phi-iqx}\Psi_{\rm cl}(p,q)
\simeq \delta\left(\phi+{2\over Q}\ln\left[2\cos\left({Q\over2}x\right)\right]\right)~,}
which is localized on the classical hairpin \hairp. 

In the opposite limit,\foot{In string theory applications $Q$ is 
actually bounded from above, $Q\le 2$, but here we are discussing boundary 
CFT where this can be ignored.} $Q\to\infty$, the wave function \onept\ goes 
like (again, neglecting an overall constant)
\eqn\onestrong{\widetilde\Psi(p,q)\simeq\Gamma(-iQp)\cos\pi{q\over Q}~,}
and the position space wave function \wavefun\ behaves like
\eqn\wavestrong{\widetilde\Psi(\phi,x)\simeq e^{-e^{-{1\over Q}\phi}}\left[\delta(x-{\pi\over Q})
+\delta(x+{\pi\over Q})\right]~.}
In this limit we have two branes localized at $x=\pm{\pi\over Q}$ and a potential
preventing the system from exploring the strong coupling region $\phi\to-\infty$. 

A more precise description of the hairpin in the limit $Q\to\infty$ is the following. 
We have a $D1$-brane and an anti $D1$-brane at a distance 
\eqn\distbranes{\delta x={2\pi\over Q}~.}
Despite the fact that the brane and anti-brane are very close, there is no instability
associated with their attraction, since for large $Q$ the translational mode of the branes in
the $x$ direction is very heavy. A condensate of the open string tachyon stretched 
between the two branes generates a boundary perturbation of the form\foot{As usual
for boundary tachyon perturbations, one has to add a boundary fermionic degree of freedom
which also appears in the boundary perturbation. We will omit it here.}
\eqn\bdrysup{\delta S_{ws}=\tilde\lambda\int d\tau d\theta e^{-{1\over 2Q}(\phi+i\tilde x)}+{\rm c.c.}~,}
where $\tilde x$ is the T-dual of $x$, $\tau$ is the coordinate along the boundary of the
worldsheet, and $\theta$ is its worldsheet superpartner. Some comments:
\item{(1)}  The interaction \bdrysup\ is the boundary $N=2$ Liouville superpotential
\eqn\bsup{W(\Phi)=\tilde\lambda e^{-{1\over 2Q}\Phi}~,\qquad \Phi=\phi+i\tilde x~.} 
It is a boundary analog of the bulk $N=2$ Liouville superpotential of \GiveonPX. 
\item{(2)} The boundary superpotential \bsup\ is non-normalizable when 
\eqn\nonnorm{-{1\over 2Q}>-{Q\over 4}~,}
\ie\ for $Q^2>2$, precisely like its bulk counterpart. 
\item{(3)} The superpotential \bsup\ leads to \wavestrong, using 
the fact that the corresponding potential behaves like 
$V=|W|^2\simeq \exp(-{1\over Q}\phi)$.

\noindent
So far we discussed the properties of the hairpin boundary state only in the limit of very small
and very large $Q$. For finite $Q$ one needs to take into account both the shape of the brane
\hairp, and the stretched tachyon condensate \bdrysup. As a concrete application of this, consider
the analytic structure of the one point function \onept. As is familiar from the cigar/Liouville 
correspondence (see \eg\ \KazakovPM) the poles of the one point function are due to
processes that can occur anywhere in the bulk of $\IR_\phi$ and can be studied by expanding 
in the interactions. 

One set of poles occurs when $iQp$ is a positive integer, and can be understood by expanding 
the one point function in powers of the boundary $N=2$ Liouville superpotential  \bdrysup. 
Winding conservation in the bulk of $\IR_\phi$ implies that non-zero contributions go like 
$|\tilde\lambda|^{2n}$ with $n$ a positive integer. Anomalous conservation of $\phi$ momentum
implies that 
\eqn\sumrule{-{Q\over2}+ip-2n{1\over 2Q}=-{Q\over2},\;\;\;{\rm or}\;\;\; iQp=n~.}
Thus, the residues of the poles at $iQp=n$ can be calculated by evaluating 
Veneziano amplitudes with $2n$ insertions of the boundary superpotential \bdrysup\
and its complex conjugate. In this calculation it is important to include
the boundary fermionic degrees of freedom that we omitted in \bdrysup.

The other set of poles of \onept\ occurs when $2ip\over Q$ is a positive integer. The residues 
of these poles can be computed by expanding the boundary action in the perturbation
\deltax\ to $n$'th order, and imposing the anomalous conservation condition
\eqn\moresum{-{Q\over2}+ip-n{Q\over 2}=-{Q\over2},\;\;\;{\rm or}\;\;\; {2ip\over Q}=n~.}
We see that in order to reproduce the analytic structure of the full amplitude \onept\ we
need to take into account both the tachyon condensate and the curved shape of the brane. 

One can also extend the analysis of \GiveonUP\ to this case and derive the one point 
function \onept\ by using degenerate operators. As there, this should give a relation 
between $\lambda$ in \deltax\ and $\tilde\lambda$ in \bdrysup. We will not perform this 
calculation (or that of the residues of the poles discussed in the previous two paragraphs) 
here. It would be nice to check that it works. 

Our main interest is in understanding the nature of $\alpha'$ corrections to the
classsical hairpin shape \hairp. To do this, we need the exact real space wave function 
of the hairpin, which is obtained by Fourier transforming the disk one point function 
\onept. This computation was done in \NakayamaYX. The result can be written in 
terms of the variable 
\eqn\defy{y={e^{-{Q\over2}\phi}\over 2\cos{Q\over2}x}}
as follows:
\eqn\exactwave{\Psi(\phi,x)\simeq 
{y^{2\over Q^2} e^{-y^{2\over Q^2}}\over 2\cos{Q\over2}x}~.}
For later use in Minkowski spacetime, it is useful to compute the
$(xx)$ component of the stress tensor associated with the brane, 
which is \NakayamaGE:
\eqn\stressxx{T_{xx}\simeq -y^{{2\over Q^2}-1}e^{-y^{2\over Q^2}}~.}
We see that the effect of $\alpha'$ corrections is to replace the classical 
solution \hairp, which has a particular value of $y$, $y=1$, with a smeared
solution, in which $y$ has a finite spread. 

The properties of the smeared solution depend on $Q$. For small $Q$, the 
stress tensor \stressxx\ is sharply peaked at  
\eqn\maxstress{y={e^{-{Q\over2}\phi}\over 2\cos{Q\over2}x}=
\left(1-{Q^2\over2}\right)^{Q^2\over2}=1-{Q^4\over 4}+\cdots~.}
As the linear dilaton slope $Q$ increases, the maximum of the distribution
moves to smaller $y$, or larger $\phi$. The distribution also becomes broader.
When $Q^2\to 2$, the peak of the distribution goes to infinite $\phi$, and its
width diverges. 

The spread of $\phi$ can be read off \defy:
\eqn\fluctphi{\delta\phi\simeq {2\over Q}\delta \ln y~.}
For small $Q$, $y\simeq 1$, \maxstress, while the width of the distribution \exactwave\ 
is $\delta y\simeq Q^2$. Hence the spread of $\phi$ is  
\eqn\ddeell{\delta\phi\sim {1\over Q}{\delta y\over y}\sim Q~.}
As $Q^2\to 2$, one finds that $\delta\phi$ diverges. 

The above properties of the boundary state can be understood in the following way. 
We saw that the main correction to the DBI picture of a brane with the shape \hairp\ 
is the condensate of the tachyon stretched between the two sides of the hairpin. For 
small $Q$ this tachyon is asymptotically very massive due to the large distance over 
which it is stretched \distbranes. Therefore, its condensate \bdrysup\ goes rapidly 
to zero as $\phi\to\infty$, and the asymptotic part of the hairpin is not significantly 
modified by its presence. As we get closer to the tip of the hairpin, the tachyon becomes 
lighter, its condensate grows and its effects increase. It makes the location of the tip 
fuzzy, and pulls it to larger $\phi$ \maxstress.

An instructive way to estimate the size of the region in which the
stretched tachyon condensate is significant, is the following. The
quadratic term in the effective action of the tachyon $T(\phi)$ has 
the form
\eqn\leff{\CL=-e^{{Q\over2}\phi}\left[(\partial_\phi T)^2+m^2T^2\right]~.}
The tachyon mass is 
\eqn\mmxx{m^2=-{1\over4}+\left(x\over2\pi\right)^2~,}
where the first term is the standard mass of the open string tachyon
in fermionic string theory, and the second is due to the stretching of
the tachyon over a distance $2x$ (at a given $\phi$). This distance 
depends on $\phi$ via the hairpin shape \hairp. As $\phi$ varies between 
infinity and its minimal value, $x$ changes between $\pi\over Q$ and zero. 

It is convenient to perform the field redefinition 
\eqn\fieldredef{T(\phi)=e^{-{Q\over4}\phi}\widetilde T(\phi)~,}
to make the kinetic term in \leff\ canonical. In terms of $\widetilde T$,
the effective Lagrangian has the form 
\eqn\lefftilde{\CL=-(\partial_\phi \widetilde T)^2-\widetilde m^2(\phi)\widetilde T^2~,}
where the effective mass $\widetilde m$ is given by
\eqn\mmtilde{\widetilde m^2(\phi)=m^2(\phi)+{Q^2\over16}~.}
As $\phi\to\infty$, the effective mass approaches the value 
\eqn\masstach{\widetilde m_\infty=\half\left({1\over Q}-{Q\over 2}\right)~.}
As we approach the tip the contribution from stretching goes to zero, and the 
effective mass goes to 
\eqn\masszero{\widetilde m_0^2=-{1\over4}+{Q^2\over 16}~.}
This value is negative for all $Q^2<4$ (\ie\ in all situations of interest except
two dimensional string theory), reflecting the fact that the open string tachyon is
tachyonic above two dimensions. 

The equation of motion of the tachyon $\widetilde T$ is
\eqn\eomtildet{-\partial_\phi^2 \widetilde T+\widetilde m^2(\phi)\widetilde T=0~.}
This looks like a Schroedinger equation for a particle living in the potential 
$V(\phi)=\widetilde m^2(\phi)$, with energy $E=0$. The analog of the Schroedinger
wave function is $\widetilde T$, which is indeed the wave function of the tachyon.
Such a wave function spreads over
the region in $\phi$ in which the potential $V(\phi)$ is smaller than the energy, 
or in other words over the region in which the effective mass squared of the 
tachyon \mmtilde\ is negative.  For small $Q$ the size of this region goes like 
$\delta\phi\sim Q$, while for $Q^2\to 2$ it diverges like 
\eqn\fluctqq{\delta\phi\sim -2\sqrt2\ln\widetilde m_\infty~,}  
in agreement with our analysis of the boundary state, in the discussion 
following eq. \fluctphi. 

We are thus led to the following picture. When the linear dilaton slope $Q$ is very 
small, the effective mass at infinity, \masstach, is very large,  and the tachyon 
wave function $\widetilde T$ decays like
\eqn\decayttilde{\widetilde T\sim e^{-\widetilde m_\infty\phi}}
for large $\phi$. As we go to smaller $\phi$, the mass squared of the tachyon
decreases and eventually becomes negative, in a region of size $\delta\phi\sim Q$
around the turning point. In that region the condensate of the tachyon is not 
suppressed. Its effect is to smear the hairpin, and shift its tip 
to slightly larger $\phi$.

As $Q$ increases, the size of the region where the tachyon is light grows, 
and the smearing fo the hairpin brane due to the light tachyon increases. 
The amount of smearing 
for $Q^2$ slightly below two is given by \fluctqq. As $Q^2\to 2$, 
$\widetilde m_\infty$ \masstach\ goes to zero, and the fluctuations extend all 
the way to infinity \fluctqq. At that point, one can no longer think about 
the boundary state in terms of the original hairpin shape.  

Earlier in this section (after eq. \hairp) we mentioned that, just like for the 
cigar geometry, the shape of the hairpin brane is not expected to receive $\alpha'$ 
corrections in the fermionic string. We now see in what sense this is correct. The 
boundary state certainly {\it does} receive $\alpha'$ corrections, which are, for 
example, visible in the form of the one point function \onept. As we discussed, 
these corrections are due to the stretched tachyon condensate \bdrysup. However, 
in position space the corrections are relatively mild: they do not change the 
asymptotic separation between the two $D1$-branes -- \distbranes\ is exact. Also, 
the hairpin equation \hairp\ is still valid in an average sense; \eg\ we saw that 
the expectation value of the stress tensor is peaked at \maxstress, which again has 
the hairpin shape \hairp, with a particular value of the constant $C$. 

\newsec{Rolling D-brane}

In this section we discuss the Minkowski continuation of the hairpin brane, which 
is obtained by replacing $x\to it$ in \hairp. This leads to the trajectory
\eqn\rollingd{e^{-{Q\over2}\phi}=2C\cosh{Q\over2}t~,}
corresponding to a $D$-brane localized in $\phi$, which starts at early times deep inside
the throat with a speed close to that of light, rises to some maximal height and then falls
back towards the strong coupling region $\phi\to-\infty$. Alternatively, one can think of
\rollingd\ for $t\ge 0$ as describing a process where the $D$-brane is released from
rest at $t=0$, and rolls down the throat. The parameter $C$ can again be set to one
by shifting $\phi$.

Unlike the Euclidean hairpin \hairp, which never makes it into the strong coupling 
region $\phi\to-\infty$, the rolling $D$-brane trajectory \rollingd\
visits this region at very early and very late times. One might worry that 
the boundary CFT describing the rolling brane will be singular, like the linear
dilaton theory itself.  A quick way to see that this is not the case \KutasovDJ\ is 
that the worldsheet CFT is typically singular when $D$-branes are light. As long 
as the energy of the rolling $D$-brane in string units is large, whether that energy 
is potential or kinetic, we do not expect the theory on the disk to exhibit any pathologies.
Indeed, the boundary state obtained by Wick rotation of the hairpin brane appears
to be well behaved.

There are potential divergences in loop amplitudes \refs{\NakayamaYX,\SahakyanCQ}, 
which are associated with  processes in which the $D$-brane loses its energy to closed 
string radiation and becomes light at late times.  We will focus on phenomena that
occur at shorter time scales, and will not discuss these quantum effects here. 

As for the hairpin, we expect the trajectory \rollingd\ to receive stringy corrections 
for finite $Q$, and our purpose is to study their nature. In Euclidean space, we saw 
that corrections to the hairpin shape \hairp\  were due to the condensate of the tachyon 
\bdrysup. A direct Wick rotation of this condensate seems subtle since the tachyon 
depends on the T-dual of $x$, which does not have a simple continuation to Minkowski 
time. Therefore, our strategy will be to examine the Minkowski boundary state, 
and use it to interpret the string corrections. 

Given the discussion of the Euclidean case, this is rather straightforward to do.
As is familiar from other contexts such as the rolling tachyon boundary state 
(see  \SenNF\ for a review), a natural way to perform the Wick rotation is in 
position space \NakayamaYX. Thus, we define a Minkowski version of the variable 
$y$ \defy, 
\eqn\defym{y_m={e^{-{Q\over2}\phi}\over 2\cosh{Q\over2}t}}
and use \stressxx\ to compute the energy density of the rolling $D$-brane, 
\NakayamaGE,
\eqn\stressoo{T_{00}=
{E\over Q\Gamma(1-{Q^2\over2})}y_m^{{2\over Q^2}-1}e^{-y_m^{2\over Q^2}}~,}
where $E$ is the total energy of the brane.

In a similar way to the Euclidean analysis of section 2, one can use \stressoo\ 
to find the location of the $D$-brane as a function of time. In the small slope 
limit $Q\to 0$ one finds that the distribution of energy density \stressoo\ is 
sharply peaked at $y_m=1$, the classical trajectory \rollingd. As $Q$ increases, 
the distribution of $y_m$ becomes wider, and its peak moves to smaller values 
\maxstress. 

The smearing of the trajectory \rollingd\ is related to that of $y_m$ as in \fluctphi,
\eqn\spp{\delta\phi\simeq {2\over Q}\delta\ln y_m~.} 
In particular, the spread of $\phi$ is independent of time, as in the Euclidean case,
where it was independent of $x$. The smearing increases with the dilaton slope, and 
pushes the top of the trajectory, $\phi(t=0)$, out into the weak coupling region. As 
in the Euclidean case, for small $Q$ one finds that the smearing is small, $\delta\phi\sim Q$,
while as $Q^2\to 2$, the top of the trajectory is pushed to infinite $\phi$, and the 
$D$-brane is completely delocalized. For larger dilaton slope, the Minkowski boundary state 
does not seem to describe the motion of a localized object in the  throat. For example, 
the integral giving the total energy,
\eqn\toten{E=\int_{-\infty}^\infty d\phi T_{00}(\phi,t)~,}
diverges for $Q^2\ge 2$. The divergence comes from the region $\phi\to\infty$.

The hairpin, \hairp, which is the Euclidean continuation of the Minkowski 
solution \rollingd\ has a periodicity in the Euclidean time direction, 
$x\sim x+{4\pi\over Q}$. This means that the natural Minkowski continuation 
corresponds to a $D$-brane in a heat bath at temperature 
\eqn\tunruh{T_u={Q\over4\pi}~.}
This temperature has a natural interpretation in terms of the dynamics of the brane. 
To see that, we recall the Unruh effect \UnruhDB, according to which an accelerated 
observer experiences a thermal bath of particles at temperature 
\eqn\unruht{T_u={a\over 2\pi}~,}
where $a$ is the acceleration of the observer. The rolling $D$-brane \rollingd\ can be 
thought of as an Unruh observer. As it starts from rest at $t=0$, its acceleration is 
given by 
\eqn\amax{a_0=-{Q\over2}~.}
The corresponding Unruh temperature \unruht\ is given by \tunruh. Thus, we see that
the heat bath in which the $D$-brane is immersed has a temperature equal to its Unruh
temperature at the top of the trajectory.

In the Euclidean case, the string corrections to the hairpin geometry were due to the
stretched tachyon \bdrysup. An interesting question, to which we turn next, is what is 
the interpretation of the corresponding corrections to the rolling $D$-brane trajectory 
\rollingd.

One way of thinking about this problem is to consider the process in which the
$D$-brane is released from rest at $t=0$, and, without taking into account 
string corrections, accelerates according to \rollingd. The Wick 
rotation from Euclidean space can be thought of as specifying an initial state 
for the rolling $D$-brane, which can be obtained by gluing the Euclidean hairpin 
solution to the Minkowski solution at $t=0$. This is an analog of the 
Hartle-Hawking prescription for black holes. 

The fact that the hairpin brane has a 
non-zero open string tachyon condensate, which is localized in the vicinity of 
the tip, suggests that the Minkowski initial state at $t=0$ contains some open 
strings. The form of the energy density \stressoo\ suggests that these open
strings oscillate in the $\phi$ direction, giving the stringy halo around the 
classical trajectory. Due to the acceleration of the $D$-brane, the open strings 
experience a force in the direction opposite the acceleration, \ie\ in the 
positive $\phi$ direction. Therefore, we expect the stringy smearing due to these 
excited strings to be skewed towards larger values of $\phi$. 

This is precisely what we see in the distribution of energy density \stressoo.
The suppression at large $y_m$ (which corresponds to $\phi\to-\infty$) is exponential, 
while that at small $y_m$ (or $\phi\to\infty$) is only powerlike, with a power that 
decreases as the acceleration \amax\ grows. Therefore, the tail of the distribution 
at large positive $\phi$ is much larger than that at large negative $\phi$.  

We are led to the following picture. For small $Q$, the excited open strings 
give rise to a small spread in $\phi$, \spp, by an amount $\delta\phi\sim Q$, 
proportional to the temperature \tunruh, and the top of the trajectory is pushed 
to slightly larger $\phi$ by the amount \maxstress. 

As $Q$ increases, the acceleration \amax\ grows, and eventually the Unruh temperature 
\tunruh\ reaches the Hagedorn temperature of perturbative string theory in the linear 
dilaton throat \rphirx, \lindil. The latter is given by (see \eg\ \GiveonMI)
\eqn\perthag{T_h={1\over 2\pi\sqrt{4-Q^2}}~.}
Thus, $T_u=T_h$ when $Q^2=2$. At that point, one expects the effect of the
classical condensate of open strings to be large. Indeed, the exact boundary 
state has the property that as $Q^2\to 2$, the trajectory develops large 
smearing, is pushed out to infinite $\phi$, and becomes non-normalizable.  

One can write the asymptotic mass of the winding tachyon \masstach, 
in the following suggestive way:
\eqn\massbeta{\widetilde m_\infty^2=
\left(1\over8\pi T_u\right)^2-\left(1\over 8\pi T_h\right)^2=
\left(\beta_u\over8\pi \right)^2-\left(\beta_h\over 8\pi\right)^2~.}
This formula is very reminiscent of the one for the mass of the thermal scalar 
in ordinary flat space string theory at finite temperature. There, it is usually 
said that the thermal scalar encodes the properties of highly excited perturbative 
strings, which dominate the critical behavior near the Hagedorn temperature. Similarly, 
in our case the condensate of the stretched tachyon \bdrysup\ in Euclidean space encodes 
the effects of the condensate of excited open strings, which for $T_u\to T_h$ consists 
mainly of very long strings. These strings cause the large smearing of the trajectory 
of the $D$-brane, which in the Euclidean case is attributed to the condensate of the 
stretched tachyon. One can describe the situation by saying that the Wick rotation of 
the stretched tachyon condensate to Minkowski space {\it is} the excited open strings.

\newsec{Black holes}

In this section we will generalize the considerations of the previous sections to the
case of black hole solutions in string theory. We start with a discussion of the two
dimensional Euclidean and Minkowski $SL(2,\IR)/U(1)$ black hole, where $\alpha'$ effects
are understood and we are on firmer ground. In subsection 4.2 we propose a generalization
to other black holes, where exact solutions are in general not available and our discussion 
will be more tentative. We will restrict to the case of uncharged black holes. It would
be interesting to generalize the discussion to charged ones.

\subsec{Two dimensional black hole}

Euclidean $SL(2,\IR)/U(1)$ CFT is a solution of the classical equations of motion of string 
theory in asymptotically linear dilaton space \refs{\ElitzurCB,\MandalTZ,\WittenYR}. The 
metric takes the cigar form 
\eqn\metcig{ds^2=d\phi^2+\left(\tanh^2{Q\over2}\phi\right) d\theta^2~,}
where $\theta$ is a periodic coordinate, 
\eqn\periodtheta{\theta\sim\theta+{4\pi\over Q}~,}
and $0\le\phi<\infty$. The dilaton is
\eqn\dil{D=D_0-\ln\cosh{Q\over2}\phi~.}
$D_0$ is the value of the dilaton at the tip of the cigar. It is a free
parameter, which should be chosen large and negative, such that the string
coupling at the tip, $\exp D_0$, is small. 

As mentioned in the introduction, in addition to the metric and dilaton 
\metcig, \dil, this CFT has a condensate of a closed string tachyon 
wrapped around the cigar \refs{\fzz,\KazakovPM,\GiveonPX}, which corresponds 
to the $N=2$ Liouville superpotential
\eqn\bulksup{W=\mu e^{-{1\over Q}\Phi}~,}
with $\Phi=\phi+i\tilde\theta$, and $\tilde\theta$ the T-dual of $\theta$.
As in the boundary case \bsup, for small $Q$ the tachyon background goes 
rapidly to zero at large $\phi$, and does not play much of a role there. 
However, as $Q$ increases, it becomes more important, until it takes over
at $Q^2=2$ and turns the cigar CFT into a non-normalizable deformation of
the linear dilaton one. 

One way to see the effects associated with the superpotential \bulksup\ is 
to study the two point function of the tachyon vertex operators \bulkt\ in 
the cigar background. It is given by \refs{\fzz,\GiveonPX,\GiveonTQ}
\eqn\twobulk{\langle T(p,q)T(-p,-q)\rangle\simeq 
{\Gamma(-{2ip\over Q})\over\Gamma({2ip\over Q})}
{\Gamma^2(\half+{ip\over Q}+{q\over Q})\over\Gamma^2(\half-{ip\over Q}+{q\over Q})}
{\Gamma(1-iQp)\over\Gamma(1+iQp)}~.}
One can think of \twobulk\ as the $1\to 1$ scattering matrix from 
the tip of the cigar and/or $N=2$ Liouville potential. As discussed in
\refs{\GiveonUP,\AharonyXN}, some of the poles of \twobulk\ can
be understood by expanding the two point function in a power series in
\bulksup, and in the metric deformation associated with \metcig\ at large
$\phi$. 

The two point function \twobulk\ has a qualitatively similar structure 
to that of the one point function \onept. The first two ratios
of $\Gamma$-functions depend on ${p\over Q}$ and ${q\over Q}$. 
This is an analog of the classical wave function \psicl. In the present case, 
one can show \DijkgraafBA\ that this is what one gets from classical geometry, 
by studying $1\to 1$ scattering of fields in the geometry \metcig, \dil. 

The last ratio of $\Gamma$-functions in \twobulk\ depends on $Qp$. 
Its effects smear the cigar geometry in the $\phi$ direction. To determine
the size of this smearing, one can proceed as in section 2.  The quadratic
term in the effective Lagrangian for the winding tachyon $T$ has the form
\eqn\actwind{\CL=\cosh^2\left({Q\over2}\phi\right)
\left[(\partial_\phi T)^2+m^2(\phi) T^2\right]~,}
where the $\phi$-dependent mass of $T$ is
\eqn\masstt{m^2(\phi)=-1+{1\over Q^2}\tanh^2{Q\over2}\phi~.}
The first term in \masstt\ is the standard mass of the closed string tachyon in 
the fermionic string, while the second is the contribution due to winding around
the $\theta$ circle. 

The wave function of the tachyon, $\widetilde T$, is obtained by dividing the 
tachyon field $T$ by the local string coupling, $g_s(\phi)$, as in \fieldredef,
\eqn\defttilde{\widetilde T(\phi)=T\cosh{Q\over2}\phi~.}
In terms of $\widetilde T$, the Lagrangian \actwind\ takes the form
\eqn\newlag{\CL=(\partial_\phi\widetilde T)^2+\widetilde m^2(\phi)\widetilde T^2~,}
where the effective mass $\widetilde m$ is
\eqn\effclmass{\widetilde m^2={Q^2\over4}-1+{1\over Q^2}\tanh^2{Q\over2}\phi~.}
As $\phi\to\infty$, it goes to 
\eqn\massbulk{\widetilde m_\infty={1\over Q}-{Q\over2}~,}
while for $\phi\to 0$ one finds
\eqn\bulkzero{\widetilde m_0^2=-1+{Q^2\over4}~.}
Note that both $\widetilde m_\infty$ and $\widetilde m_0$ are 
twice as large as the corresponding open string quantities \masstach\
and \masszero. In particular, $\widetilde m_0^2$ is negative.

The equation of motion of $\widetilde T$ is again given by \eomtildet, and looks
like the Schroedinger equation for a zero energy wave function in the potential
$\widetilde m^2(\phi)$. The wave function spreads over the region $0\le\phi\le \phi_0$, 
in which $\widetilde m^2(\phi)$ is negative. From \effclmass\ we see that
$\phi_0$ is given by 
\eqn\phinot{{1\over Q^2}\tanh^2{Q\over2}\phi_0=1-{Q^2\over4}~.}
For small $Q$, one has $\phi_0=2+O(Q^2)$. Note that this is different
from the open string case, where the spread of the hairpin for small $Q$
was by an amount proportional to $Q$, \ddeell. 
As $Q^2\to 2$, $\phi_0$ diverges in a way similar to \fluctqq,
\eqn\divphio{\phi_0\sim -\sqrt{2}\ln\widetilde m_\infty~.}
Thus, we are led to the following picture. For a fixed value of 
$Q<\sqrt2$, far from the tip of the cigar, the $SL(2,\IR)/U(1)$ CFT 
is well described by the metric and dilaton \metcig, \dil. However,
in a region of size $\phi_0$, \phinot, there are large fluctuations of the 
geometry, which are due to the presence of a light winding tachyon 
condensate \bulksup. The size of this region grows with $Q$, and 
diverges as $Q^2\to 2$, \divphio, where the cigar CFT ceases to be 
normalizable. We see that the picture is very similar to the one 
obtained for the hairpin brane in section 2. 

The Minkowski two dimensional black hole is obtained by Wick rotating
$\theta\to it$ in \metcig. It can also be studied algebraically by using 
its $SL(2,\IR)/U(1)$ description. The Hawking temperature of this black 
hole is 
\eqn\bekaw{T_{bh}={Q\over 4\pi}~.}
Note that \bekaw\ is equal to the Unruh temperature of the rolling $D$-brane
discussed in section 3, \tunruh. $T_{bh}$ also has the interesting property
that it is independent of the mass of the black hole. This too has
an analog in the rolling $D$-brane problem, for which the Unruh
temperature \tunruh\ is independent of the energy of the $D$-brane. 

Due to the close analogy to the rolling $D$-brane, we can borrow
most of the discussion of section 3 to the present case. The condensate
of the winding tachyon in the Euclidean black hole solution \metcig\
corresponds in Minkowski space to a condensate of closed strings near
the horizon of the black hole. This leads to the smearing of the region
near the horizon, and the appearance of a stretched horizon of the sort
discussed in \SusskindWS. The size of this stretched horizon is given by 
\phinot. 

This size has a very natural Minkowski interpretation. While the Hawking 
temperature seen by an observer at infinity is given by \bekaw, an observer 
stationary at a fixed value of $\phi$ sees a higher temperature \UnruhIC, 
due to the red-shift factor in \metcig. That temperature is given by  
\eqn\localhawk{T_{bh}(\phi)={Q\over4\pi}{1\over \tanh{Q\over2}\phi}~.}
Even if the Hawking temperature at infinity \bekaw\ is low, the local temperature
\localhawk\ increases as one approaches the horizon, where it formally diverges.
It is natural to expect that stringy effects will smear the part of the geometry for
which the local Hawking temperature is higher than the Hagedorn one \SusskindWS.
Comparing \localhawk\ to \perthag\ we see that the place where the two coincide
is precisely \phinot. As we saw before, the size of the stretched horizon is of
order one in string units for black holes with very low Hawking temperature, and
it diverges as the Hawking temperature at infinity approaches the Hagedorn temperature.

In the rolling $D$-brane case we described the situation by saying that the Minkowski
continuation of the stretched tachyon condensate \bsup\ is the classical condensate 
of open strings that smears the trajectory of the $D$-brane. Similarly, we can 
say that the continuation of the winding tachyon \bulksup\ is the classical condensate
of closed strings near the horizon, which smears the geometry in a region of size \phinot.
The mass of the wound tachyon \massbulk\ can be written in a way analogous to \massbeta,
\eqn\massblack{\widetilde m_\infty^2=\left(1\over 4\pi T_{bh}\right)^2-
\left(1\over 4\pi T_h\right)^2=\left(\beta_{bh}\over 4\pi \right)^2-
\left(\beta_h\over 4\pi \right)^2~,}
which is also suggestive of this interpretation, as in the discussion after \massbeta.

The notion that a small black hole has a stretched horizon whose size is much
larger than the horizon size sounds counter-intuitive. The following remarks
may help clarify this issue.

In a black hole spacetime, one can divide the modes of quantum fields into two 
classes. One consists of modes that originate far from the black hole; these are 
refered to in \UnruhIC\ as ``past-infinity'' modes. The other consists of modes 
that appear to an outside observer to emanate from the black hole. These are the 
so called ``white hole'' modes. For a black hole in vacuum, the white hole modes
are in a thermal state at their Hawking temperature \bekaw, \localhawk, but the
past infinity modes are in their vacuum state. On the other hand, for a black hole
in thermal equilibirium with a heat bath, both kinds of modes are in a thermal
state at the relevant temperature. 

An observer at the distance \phinot\ from a two dimensional black hole will
perceive thermal radiation at the Hagedorn temperature coming out of the black hole,
but depending on the presence or absence of a heat bath at the Hawking temperature,
will or will not feel thermal radiation travelling towards the hole. In $1+1$ dimensions, 
the difference between the two situations seems to be small. 
Thus, in this case, an observer at a fixed $\phi$ will indeed perceive the 
local Hawking temperature \localhawk, even if the radiation from infinity is absent.  

In higher dimensions, the case discussed in the next subsection, the difference
is bigger (for small black holes), since the radiation in white hole modes disperses
over the angular spheres. If the thermal bath is absent, an observer at a given radial
distance may see very faint radiation at close to the Hagedorn temperature, but will
not feel that the local temperature around him is $T_h$. The continuation from Euclidean 
space naturally describes the black hole in thermal equilibrium with a heat bath, and in
that system, one can say that the effective size of the black hole diverges when the 
Hawking temperature approaches the Hagedorn one.

\subsec{Higher dimensional black holes}

In the previous subsection we identified the mechanism for the appearance
of a stretched horizon for classical two dimensional black holes. According 
to \SusskindWS, a similar stretched horizon is expected to surround general 
black holes in string theory. It is natural to ask whether our arguments can 
be generalized to other types of black holes, such as Schwarzschild. In this 
subsection we discuss this question.

Consider, as an example, a Schwarzschild black hole in asymptotically flat four 
dimensional spacetime.\foot{With the six extra dimensions compactified.} We 
will assume that the classical equations of motion of string theory have a solution 
which looks like an $\alpha'$ corrected Schwarzschild solution. In other words, 
we are assuming the existence of a one parameter set of solutions labeled by 
the mass, which for large masses approaches the Schwarzschild solution, but 
also makes sense for masses for which the Hawking temperature is of order the 
Hagedorn temperature. The basic properties we require of such a solution are 
spherical symmetry, the existence of a horizon, and of a Wick rotated Euclidean 
solution, in which Euclidean time is compactified on a circle with asymptotic 
circumference $\beta_{bh}=1/T_{bh}$. The circle should be contractible, \ie\
the winding number around it should not be conserved.

It is possible that this assumption is incorrect, but we do not know of any reason
why this should be the case, and the two dimensional black hole provides an existence 
proof that such solutions are possible. It is very natural for such solutions to exist,
since the masses for which the Hawking temperature is of order $T_h$ go like
$1/g_s^2$, and spherically symmetric states with such masses should be described in the 
limit $g_s\to 0$ by non-trivial closed string geometries. Moreover, such solutions should
have horizons since they should give rise to the large entropy of states with these masses.
In any case, we will assume that they exist and proceed.

The Hawking temperature of the black hole, $T_{bh}$, decreases as its mass increases, 
and for large enough mass can be made arbitrarily small compared to the Hagedorn 
temperature. However, this is the temperature at infinity. An observer at a fixed 
proper distance $R$ from the horizon sees a higher Hawking temperature,
\eqn\localhhh{T_{bh}(R)={T_{bh}\over \chi(R)}~,}
where $\chi$ is the redshift factor in the black hole metric, $\chi=\sqrt{g_{00}}$.
For large $R$, $\chi\to 1$, while as $R$ decreases, $\chi$ does as well, so the
local Hawking temperature \localhhh\ increases. In particular, as $R\to 0$, the 
local Hawking temperature formally diverges. For small $R$, $T_{bh}(R)$ is more 
naturally thought of as an Unruh temperature \unruht, due to the acceleration of an 
observer held at a fixed distance from the horizon; see \eg\ \UnruhIC\ for a discussion. 

When the local Hawking temperature \localhhh\ approaches the Hagedorn temperature 
in flat spacetime,\foot{Which is given by the limit $Q\to 0$ of \perthag.} 
\eqn\hagflat{T_h={1\over 4\pi}~,}
one expects stringy effects to become important. These effects are expected to
smear the geometry into a stretched horizon \SusskindWS, which can be defined 
as the region where 
\eqn\defstretch{T_{bh}(R)\ge T_h~.} 
For a large black hole, the size of the stretched horizon is of order one in string 
units (see below). 

As the mass of the black hole decreases, and the corresponding Hawking 
temperature increases, we encounter a conundrum. On the one
hand, as the mass decreases, the gravity fields decay more rapidly at infinity.
In particular, the Einstein gravity approximation to the full solution becomes
better and better for any fixed radial distance from the black hole, and the
geometry should look more and more like flat space there. On the other
hand, as the Hawking temperature increases, we expect the stretched horizon
defined via \defstretch\ to increase in size, since a smaller redshift factor
is needed to bridge the gap between the temperature at infinity, $T_{bh}$, 
and the Hagedorn temperature, $T_h$. If the gravity solution becomes more 
and more like flat space, what causes the smearing of the geometry at long
distances that is implied by a large stretched horizon? 

The discussion of the previous subsection suggests an obvious answer. The 
Euclidean Schwarzschild geometry contains not just the gravity fields that are 
usually considered, but also a condensate of a tachyon wrapped around the 
Euclidean time circle. For very massive black holes, this tachyon is asymptotically 
very heavy, due to the large radius of Euclidean time, and so its effects at infinity 
are very small. However, as in our discussion of the hairpin brane in section 2, and 
of the cigar in the previous subsection, there is a small region near the horizon in 
which this tachyon is light, and its effects lead to a smearing of the geometry there.

As the Hawking temperature increases, the asymptotic mass of the tachyon decreases,
and the size of the stretched horizon grows, until for $T_{bh}\to T_h$ it diverges,
and the black hole ceases to be normalizable. 

To illustrate the above ideas, we will generalize the calculations of the previous
subsection to the four dimensional Schwarzschild black hole case. In order
to study the geometry at the string scale, we need the exact solution of classical 
string theory corresponding to this black hole, which is at present unavailable.
In the spirit of \HorowitzNW, we will use the leading order solution (\ie\ the original
Schwarzschild one), and hope that the $\alpha'$ corrections do not modify the
discussion too much. Of course, for any black hole for which the exact solution
is available, we can do better.

The line element of the four dimensional Euclidean Schwarzschild solution is 
\eqn\fourdsch{ds^2=\left(1-{r_0\over r}\right)d\theta^2+
{1\over 1-{r_0\over r}}dr^2+r^2d\Omega~,}
where $r_0$ is the Schwarzschild radius, and $\theta$ is the Euclidean time 
coordinate. It is periodically identified $\theta\sim \theta+\beta_{bh}$, where 
$\beta_{bh}$ is the inverse Hawking temperature. The different parameters
are determined by the mass of the black hole as follows:
\eqn\detpar{\eqalign{r_0=&{2MG_N}~,\cr
\beta_{bh}=&4\pi r_0=8\pi M G_N~,\cr
}}
where $G_N$ is Newton's constant. The inverse Hagedorn temperature of string theory 
in flat spacetime is given by $\beta_h=4\pi$, \hagflat. Hence, if we trust the 
leading order solution \fourdsch, the Hawking temperature \detpar\ is lower 
than the Hagedorn one for $r_0>1$. 

The boundary conditions around the $\theta$ circle at $r\to\infty$ are such 
that  a tachyon wrapped once around the circle survives the GSO projection. 
As discussed above, it is natural to expect that this tachyon has a non-zero 
expectation value in the solution corresponding to \fourdsch\ in string theory. 
To understand where in $r$ that expectation value is supported, we examine
the quadratic terms in the action for the tachyon. Taking the tachyon to be an 
s-wave on the angular two sphere, the action takes the form (up to an overall
constant, which will not play a role)
\eqn\actschtach{S=\int_{r_0}^\infty dr r^2
\left[\left(1-{r_0\over r}\right)(\partial_r T)^2+m^2(r) T^2\right]~.}
The mass $m(r)$ has again two contributions, the usual closed string
tachyon mass and a term from winding around the circle:
\eqn\mmrr{m^2(r)=-1+r_0^2\left(1-{r_0\over r}\right)~.}
As $r\to\infty$, the mass approaches  
\eqn\mminfty{m_\infty^2=r_0^2-1~. }
Since we are
assuming that the asymptotic Hawking temperature \detpar\ is below the
Hagedorn temperature \hagflat, $m_\infty^2$ is positive. As $r\to\infty$,
$T$ behaves as a standard scalar field with mass \mminfty, and its 
condensate decays like 
\eqn\condasy{T(r)\sim {1\over r}e^{-m_\infty r}~.}
On the other hand, as $r\to r_0$, the mass \mmrr\ goes to $m^2\to -1$. 
The tachyon condensate is a solution of the equation of motion of \actschtach,
and we expect it to have some finite spread in the $r$ direction. To find this
spread it is convenient to first change coordinates to 
\eqn\chvar{e^z=1-{r_0\over r}~.}
In the new coordinates, the horizon corresponds to $z=-\infty$, while $r=\infty$
is mapped to $z=0$. 

In these coordinates, the action \actschtach\ takes the form 
\eqn\newact{S=r_0\int_{-\infty}^0dz\left[(\partial_z T)^2+\widetilde m^2(z) T^2\right]~,}
where
\eqn\ttllmm{\widetilde m^2(z)=r_0^2{e^z\over (1-e^z)^4}(r_0^2e^z-1)~.}
As in the other cases described above, the tachyon wave function spreads over the
region in which $\widetilde m^2$ is negative,
\eqn\spreadz{r_0^2 e^z<1~.}
In terms of the original $r$ coordinates \chvar, this region is 
\eqn\spreadrr{1-{r_0\over r}<{1\over r_0^2}~.}
By using \localhhh, \hagflat, \detpar, one can check that \spreadrr\
is precisely the region \defstretch, in which the local Hawking temperature
is higher than the Hagedorn temperature.

If the black hole is very massive, such that $r_0\gg 1$, 
it is more natural to express \spreadrr\ in terms of the proper
distance from the horizon, $R$. For $r-r_0\ll r_0$, the 
proper distance is given by 
\eqn\propdist{R=2r_0\sqrt{1-{r_0\over r}}~.}
Hence, we can write \spreadrr\ as $R<2$. We see that for large
black holes the stretched horizon has size of order one in string units.
In the other extreme limit, when the Hawking temperature of the black 
hole approaches the Hagedorn temperature, \ie\ $r_0\to 1$, the size
of the stretched horizon diverges. Indeed, in that limit  \spreadrr\ 
takes the form 
\eqn\largerr{r<{1\over 2(r_0-1)}~.}
Note that while the detailed estimate of the size of the stretched horizon for 
very massive black holes relies on the structure of the (unknown) $\alpha'$ 
corrections to the solution \fourdsch, the estimate \largerr\ does not. Indeed, 
in the limit $r_0\to 1$ the tachyon mass \mminfty\ goes to zero, so the effective 
action \actschtach\ becomes more and more accurate. Furthermore, at the 
distance scale \largerr, the metric is essentially flat, so $\alpha'$ corrections 
to the solution should not play a role. 

To find the energy at which the size of the stretched horizon diverges,
we need to know the relation between the mass and  $r_0$ which generalizes
\detpar. This does require knowledge of the $\alpha'$ correction. If we knew
that relation, we would be able to calculate the precise energy at which the
black hole becomes non-normalizable. This should also be the point at which
the string/black hole transition occurs, since when the Hawking and Hagedorn
temperatures approach each other, it becomes more and more difficult to 
distinguish the black hole with its infinitely large stretched horizon from 
a string state with the same energy. Therefore, at that point, the 
Bekenstein-Hawking and string entropies must agree. 

To get a feeling for how this works, imagine that the Schwarzschild geometry
\fourdsch\ received no $\alpha'$ corrections (an assumption that is surely
incorrect). Then we would proceed as follows. We saw that the Hawking
and Hagedorn temperatures coincide when the horizon size $r_0=1$. 
At that point, the Bekenstein-Hawking entropy of the black hole is given
by
\eqn\bekhawk{S_{bh}={A\over 4G_N}={\pi r_0^2\over G_N}={\pi\over G_N}~.}
The mass of the black hole is given by \detpar\ 
\eqn\masssmall{M={1\over2G_N}~.}
Substituting this into \bekhawk\ we see that the Bekenstein-Hawking
entropy at the transition point is given by 
\eqn\enttran{S_{bh}=2\pi M~.}
On the other hand, the fundamental string entropy is given by \hagflat\
\eqn\entpert{S_f=4\pi M~.}
We see that the two differ by a factor of two,\foot{For a $d$ dimensional
Schwarzschild black hole, this factor is $(d-2)/(d-3)$.} which is very 
likely due to higher $\alpha'$ corrections. One can try to study this
systematically by including (tree level) higher curvature corrections
to the Einstein equations, but we will not attempt to do this here.

\newsec{Discussion}

In this paper we studied the $\alpha'$ effects in some Euclidean and Lorentzian 
backgrounds of classical string theory. One of the solutions we discussed 
corresponds to a $D$-brane rolling in a linear dilaton space. We argued that 
the main string correction to the trajectory of such a $D$-brane is a stringy
smearing due to the presence of oscillating open strings attached to the brane.
These effects increase with the gradient of the dilaton, and diverge at a critical
value of the gradient, at which the Unruh temperature of the brane reaches the
Hagedorn temperature of perturbative string theory in the linear dilaton background.
At that point the trajectory of the $D$-brane is completely smeared, and the brane
is pushed out of the linear dilaton throat. 

The Euclidean solution corresponding to the rolling $D$-brane (the hairpin brane) 
encodes these effects via the condensate of an open string tachyon stretched in 
the Euclidean time direction. In Euclidean space this tachyon smears the hairpin 
brane by an amount whose size increases with the gradient and 
diverges when the mass of the open string tachyon at infinity goes to zero. 

We also discussed Euclidean and Lorentzian black hole solutions. We argued that in this 
case $\alpha'$ corrections give rise to the stretched horizon of \SusskindWS, \ie\ they 
smear the geometry in a region outside the horizon of the black hole in which the 
locally measured Hawking temperature exceeds the Hagedorn temperature. When the 
asymptotic Hawking temperature of the black hole is small, the size of the stretched 
horizon is of order one in string units, but as it increases, the stretched horizon 
grows and eventually extends all the way to infinity when the Hawking temperature 
reaches the Hagedorn one. 

The stretched horizon is due to a condensate of closed strings near the horizon of the
black hole. In Euclidean space, this is encoded by a condensate of a tachyon wrapped 
around the Euclidean time direction. This condensate smears the Euclidean black hole 
geometry. The size of the smearing increases with the Hawking temperature of the black 
hole, and diverges when the mass of the wound tachyon at infinity goes to zero.

The above description of black holes with Hawking temperature of order the
string scale is relevant for the string/black hole transition 
\refs{\SusskindWS,\HorowitzNW}. A puzzle that was raised in these
papers, and was further discussed in \refs{\HorowitzJC,\DamourAW} is the
following. Near the transition, the size of the black hole, as measured by
its Schwarzschild radius, is of order one in string units. However, for small
string coupling, a generic string state with the same energy is very large. 
How could a small black hole smoothly turn into a large string? 

It was proposed in \refs{\SusskindWS,\HorowitzNW,\HorowitzJC,\DamourAW}
that one ingredient in the resolution of the puzzle is that string states might shrink 
due to the effects of gravitational self-interactions. Here we discussed another 
relevant effect. As the Schwarzschild radius decreases and the Hawking
temperature increases, the size of the stretched horizon grows, until the Hawking
temperature reaches the Hagedorn temperature, where it diverges. Thus, for a
Hawking temperature slightly below the Hagedorn temperature, a classical black hole
is surrounded by a very large stringy halo, and becomes harder and
harder to distinguish from a generic free string state with the same energy. 
This picture explains why the entropies of strings and black holes must agree
at the transition point. 

As mentioned in section 4, the above picture is strictly valid for a black hole
in a heat bath at its Hawking temperature, which is what one gets by continuation
from Euclidean space. In this case, an observer at a fixed distance from the 
horizon perceives a thermal bath at the local Hawking temperature \localhhh.
When the Hawking temperature at infinity approaches the Hagedorn temperature,
the properties of the system are dominated by the heat bath, and it becomes more 
and more difficult to distinguish it from fundamental strings at the same temperature. 

When the heat bath is absent, the ``white hole''  modes are still
in a thermal state at the local Hawking temperature, but the ``past infinity'' modes
are in their ground state. In this situation, it might be argued that the definition 
\defstretch\ of the stretched horizon is less natural, since an observer at that distance
does not perceive the temperature around him to be $T_h$. However, it is possible 
that in that case too, the definition \defstretch\ is meaningful: for shorter distances the
black hole can not be distinguished from a fundamental string state, while for much 
larger distances it can.

The above discussion also provides in principle a way to compare the entropies of black
holes and strings, including coefficients. If one can construct a $d$ dimensional Euclidean 
Schwarzschild black hole in asymptotically flat spacetime as an exact worldsheet  CFT, we
expect that solution to have the following properties:
\item{(1)} The closed string tachyon with winding number one around Euclidean
time at infinity should have a non-zero condensate. This condensate is normalizable 
when the Hawking temperature is below the Hagedorn temperature (\ie\ the 
circumference of Euclidean time at infinity is larger than $4\pi$) and non-normalizable 
otherwise. Thus, the Euclidean Schwarzschild solution should be non-normalizable 
when the Hawking temperature is above the Hagedorn temperature.
\item{(2)} At the point where the Hawking temperature of the $\alpha'$ corrected
solution is equal to the Hagedorn temperature of strings in flat spacetime, \hagflat, 
the black hole entropy should agree with the string one, \entpert.  

\noindent
For  two dimensional black holes for which the exact solution is known, the above
properties were verified in \GiveonMI. For $d$ dimensional Schwarzschild black holes, we saw 
in section 4 that even without taking  $\alpha'$ corrections into account, the black hole
and string entropies agree to within a factor $(d-3)/(d-2)$. It would be interesting to see
whether the disagreement decreases when one takes into account higher order 
$\alpha'$ corrections to the Einstein Lagrangian.

Our discussion is also related to that of \AtickSI, where it was argued that 
string theory at finite temperature undergoes a first order phase transition,
at a temperature below the Hagedorn temperature, which is described in 
the Euclidean time formalism by condensation of the thermal scalar, a 
tachyon which winds once around Euclidean time. Physically, one would
expect the transition to be to a Euclidean black hole, which as we saw is indeed 
likely to have a non-zero condensate of the wound tachyon, and moreover,
slightly below the Hagedorn temperature, this wound tachyon is the
leading deviation from flat space that one sees at infinity (since the horizon of
the black hole has size of order $l_s$ at that point). See \BarbonDD\ and
references therein for some related comments.

The phenomena discussed in this paper have other applications as well.
For example, it was shown in \refs{\FelderXU,\SaremiYD} that the DBI
Lagrangian describing the dynamics of $D$-branes in the gravitational
potential of other objects, such as $NS5$-branes or other $D$-branes,
has the property that generic solutions of the equations of motion develop
inhomogeneities that grow with time, and caustics. Our analysis of the 
hairpin brane shows that when such inhomogeneities start forming, the 
curved regions can become fuzzy in the same way as the vicinity of the tip 
of the hairpin in our discussion in section 2. This occurs since such 
curved branes have non-zero condensate of the open string tachyon, which
is light near local minima of the $D$-brane shape, and its fluctuations
smear the shape of the brane there. Thus, $\alpha'$ effects are expected
to smooth out the singularities seen in the DBI approximation.  Similarly, 
more general solutions corresponding to accelerating $D$-branes in string 
theory are expected to develop stringy smearing similar to that found in 
section 3.

\bigskip
\noindent{\bf Acknowledgements:}
I am grateful to T. Banks, T. Damour, A. Giveon, N. Itzhaki, E. Martinec, 
A. Parnachev, J. Polchinski, E. Rabinovici, D. Sahakyan, A. Strominger 
and A. Zamolodchikov for useful discussions.  I also thank the Aspen 
Center for Physics and Rutgers NHETC for hospitality. 
This work was supported in part by DOE grant \#DE-FG02-90ER40560.

\listrefs
\end